\documentclass[pdflatex,sn-mathphys-num]{sn-jnl}

\usepackage{graphicx}%
\usepackage{multirow}%
\usepackage{amsmath,amssymb,amsfonts}%
\usepackage{amsthm}%
\usepackage{mathrsfs}%
\usepackage[title]{appendix}%
\usepackage{xcolor}%
\usepackage{textcomp}%
\usepackage{manyfoot}%
\usepackage{booktabs}%
\usepackage{algorithm}%
\usepackage{algorithmicx}%
\usepackage{algpseudocode}%
\usepackage{listings}%

\newcommand{\Tr}{\textrm{Tr}}

\newcommand{\be}{\begin{equation}}
\newcommand{\ee}{\end{equation}}

\newcommand{\lf}{\left}
\newcommand{\rg}{\right}
\newcommand{\ra}{\rangle}
\newcommand{\la}{\langle}

\newcommand{\bea}{\begin{eqnarray}}
\newcommand{\eea}{\end{eqnarray}}

\newcommand{\cyt}{\mathcal{C}_{\mathcal{T}}}
\newcommand{\cyr}{\mathcal{C}_{\mathcal{R}}}

\usepackage{xcolor}
\usepackage{color}
\usepackage{hyperref}
\hypersetup{colorlinks=true,citecolor={blue},linkcolor={blue},urlcolor={blue}}

\theoremstyle{thmstyleone}%
\usepackage{array}
\theoremstyle{thmstyletwo}%

\theoremstyle{thmstylethree}%

\begin{document}


\title[Thouless quantum walks in topological flat bands]{Thouless quantum walks in topological flat bands}


\author[1]{\fnm{Carlo} \sur{Danieli}}

\author*[1,2]{\fnm{Laura} \sur{Pilozzi}}\email{laura.pilozzi@cnr.it}

\author[3,2]{\fnm{Claudio} \sur{Conti}}

\author[1]{\fnm{Valentina} \sur{Brosco}}

\affil*[1]{\orgdiv{Institute for Complex Systems}, \orgname{National Research Council (ISC-CNR)}, \orgaddress{\street{Via dei Taurini 19}, \city{Rome}, \postcode{00185}, \country{Italy}}}

\affil[2]{\orgname{Research Center Enrico Fermi}, \orgaddress{\street{Via Panisperna 89a}, \city{Rome}, \postcode{00184}, \country{Italy}}}

\affil[3]{\orgdiv{Department of Physics}, \orgname{University of Sapienza}, \orgaddress{\street{Piazzale Aldo Moro 5}, \city{Rome}, \postcode{00185}, \country{Italy}}}

\abstract{Non-Abelian gauge symmetries are cornerstones of modern theoretical physics, underlying fundamental interactions and the geometric structure of quantum mechanics. However, their potential to control quantum coherence, entanglement, and transport in engineered quantum systems remains to a large extent unexplored. In this work, we propose utilizing non-Abelian Thouless pumping to realize one-dimensional discrete-time quantum walks on topological lattices characterized by degenerate flat bands. Through carefully designed pumping cycles, we implement different classes of  holonomic coin and shift operators. This framework allows for the construction of quantum walks that encode the topological and geometric properties of the underlying system. Remarkably, the resulting evolution exhibits parity symmetry breaking and gives rise to a dynamical process governed by a Weyl-like equation, highlighting the deep connection between parity and time-reversal symmetry breaking in the system.}


\keywords{Thouless pumping, quantum computing, quantum walks }

\maketitle

\section{Introduction}\label{sec1}

Quantum walks (QWs) describe the quantum evolution of a system on a graph~\cite{PhysRevA.48.1687,PhysRevLett.102.180501,Kempe,aharonov2002quantum}. As a fundamental concept in quantum information science and quantum kinetics, QWs have applications in network exploration, quantum information processing, and even biochemical systems~\cite{childs2004,berry2011,sanchezburillo2012}. They provide a natural framework for studying quantum transport and diffusion,  integrating concepts from graph theory, Markov processes, and quantum mechanics~\cite{qiang2024}.
Originally introduced as the quantum counterparts of classical random walks, QWs exploit quantum interference and entanglement to shape the final statistical distribution and extend the walker's path length. Their computational universality~\cite{PhysRevLett.102.180501,lovett2010universal} makes them instrumental in the development of quantum algorithms~\cite{farhi1998quantum,Kempe,qiang2024,szegedy2004spectra,Grover1996AFQ,pub.1089200367}, with applications in quantum search~\cite{PhysRevA.67.052307,PhysRevA.82.042333}, simulation~\cite{PhysRevLett.108.010502}, and quantum communication~\cite{Srikara2020quantum}.

QWs take place on graphs $\mathcal{G = (V, E)}$, where the vertices $\mathcal{V}$ represent the sites the walker can visit, and the edges $\mathcal{E}$ define the connections between the sites.
While space in QWs is discrete \cite{Kempe}, time can be either discrete or continuous -- yielding continuous-time~\cite{PhysRevLett.102.180501,farhi1998quantum} and discrete-time~\cite{nayak2000quantum} QWs. 
The state of the walker is described by its position on the graph and a  quantum coin variable, such as spin or photon polarization. In the \textcolor{black}{Discrete-Time Quantum Walk (DTQWs)}, the walker evolves through a sequence of unitary maps, each consisting of a coin operator $\mathcal{R}$, which acts on the coin state, and a conditional shift operator $\mathcal{T}$, which moves the walker depending on the coin state. The Hadamard is a common coin operator \cite{aharonov2002quantum}, however,  DTQWs can be defined with different parametric families of quantum coin operators for better control and optimization \cite{Tregenna2003controlling,Chandrashekar2008optimizing}.

\textcolor{black}{By appropriate choice of the coin operation and underlying graph topology, one can tailor the walk dynamics to implement problem-specific symmetries, introduce controlled disorder, or explore symmetry-breaking phenomena. Suitably designed DTQWs can simulate arbitrary quantum circuits, thereby achieving computational universality \cite{lovett2010universal}. This universality is conditional in the sense that it depends on the specific configuration of the walk. The underlying graph must be appropriately engineered and the set of available coin operators must be sufficiently rich to generate a universal gate set. Under these constraints, the DTQWs formalism provides a complete and physically implementable model of quantum computation, equivalent in power to the standard circuit model. }

In this work, we propose a general approach to construct \textcolor{black}{DTQWs} using non-Abelian holonomies. Holonomic transformations enable full control of the state of quantum systems and allow for the generation of topologically quantized transport on lattices~\cite{Thouless1983quantization}. We show that these two ingredients can be combined to engineer quantum walks with tunable features and investigate their inherent relation with non-Abelian Thouless pumping~\cite{PhysRevA.103.063518,Pilozzi2022,Citro2023thouless}.

Thouless pumping is a paradigmatic topological phenomenon, yielding quantized transport in slowly and cyclically modulated one-dimensional lattices~\cite{Thouless1983quantization}.  In the presence of degenerate Bloch bands, Thouless pumping acquires a non-Abelian character~\cite{PhysRevA.103.063518}. In this case, the system is initially prepared in a Wannier state belonging to a degenerate band and undergoes a  geometric evolution dictated by the Wilczek–Zee connection~\cite{PhysRevLett.52.2111}.

Here, we employ non-Abelian Thouless pumping to generate discrete-time quantum walks. To keep the discussion simple, we illustrate our results on a Lieb chain with two degenerate flat bands~\cite{leykam2018artificial,leykam2018perspective,vicencio2021photonic,danieli2024flat}. However, our approach  can be extended in multiple directions---by considering alternative internal degrees of freedom for encoding the coin state, or by implementing lattices with different topologies and dimensionalities.

We show that holonomic transformations allow the realization of arbitrary unitary coin operations---going beyond the standard Hadamard coin---and unidirectional, conditional shift operators. We refer to this new class of one-dimensional quantum walks as {\it Thouless holonomic Quantum Walks} (ThQWs). The structure of ThQWs is intrinsically linked to the topological and  geometric properties of the Hilbert space and can be engineered by suitably designing the elementary pumping cycles.

We show that ThQWs  enable the selective breaking of parity or time-reversal symmetry, allowing the engineering of distinct final quantum-correlated states.
The parity-breaking nature of ThQWs is further highlighted by relating the dynamical equation governing the walk to the Weyl equation. We focus on a possible implementation in photonic waveguide arrays~\cite{PhysRevA.103.063518,Sun2022}, where the propagation coordinate $z$ plays the role of time. However, our results are general and can be extended to other platforms, including cold atoms in optical lattices~\cite{danieli2024nonabelian} and superconducting nanocircuits~\cite{brosco2008}.

\section{Results}\label{sec2}

Discrete time quantum walks are discrete processes that take place on graph a $\mathcal{G}$ consisting of vertices $\mathcal{V}$ and edges $\mathcal{E}$ connecting them. Each vertex is associated with a set of quantum coin variables $\sigma$.  
The wavefunction at time $t\in \mathbb{Z}$ is written as $|\Psi(t)\ra = \sum_{\bf n} \sum_\sigma   \psi_n^\sigma (t) |{\bf n},\sigma \rangle$, where ${\bf n}$ labels the vertex $v\in\mathcal{V}$. 
A single time-step, advancing from $t$ to $t+1$ consists of two unitary operations: a coin operator $\mathcal{R}$ that couples the coin states within each vertex of the graph, and a conditional shift operator $\mathcal{T}$ that moves the walker along the edges conditionally to its coin level $\sigma$.  
Although DTQWs have been  realized in various platforms~\cite{ryan2005experimental,giordani2019experimental,Zhu2020photonics,gong2021,flurin2017,PhysRevLett.120.260501}, their experimental implementation still poses  significant technical challenges related to preserving discreteness in both time and space.

\begin{figure}[h]
	\centering
	\includegraphics [width=0.5\columnwidth]{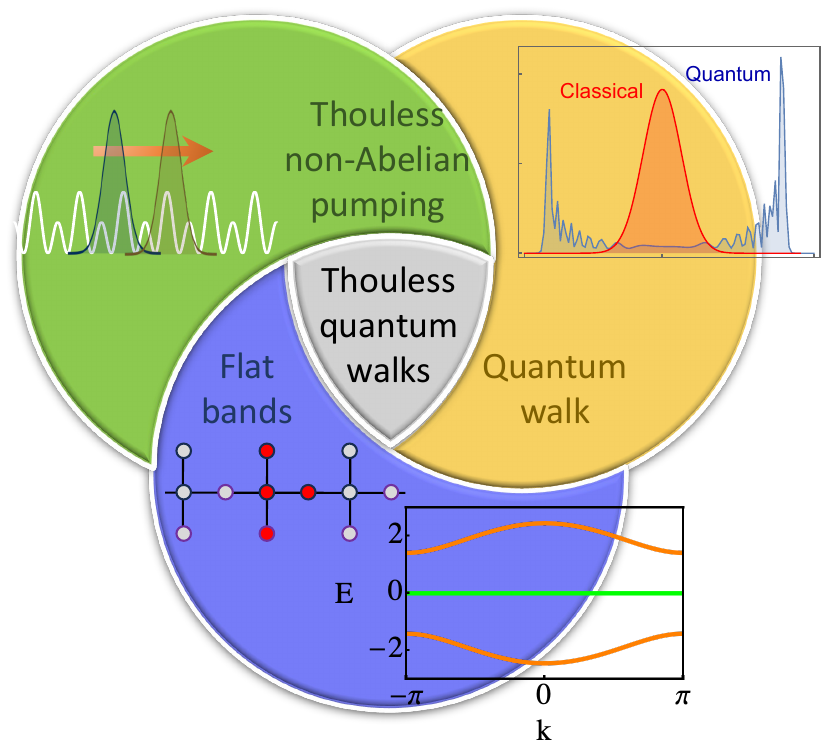}
	\caption{
	{\bf Inception of Thouless quantum walk.} 
	Essential elements underlying the construction of Thouless quantum walk. 
	In orange the discrete time quantum walk, whose compositions of coin and shift operators upon the level states induce quantum superposition that allow to overcome the mean walker’s path length beyond classical random walks. 
	In blue flat band lattices, whose destructive interference grants the existence of degenerate orthogonal spatially compact states associated to the non-dispersive bands which mimic the coin levels of a quantum walk. 
	In green non-Abelian Thouless pumping, whose cycles grant both conditional transport and geometrical unitary superposition of coin levels mimicking  coin and shift operators of a quantum walk. 
	}
	\label{fig:intro}
\end{figure}

In this work, we introduce ThQWs as an elegant and robust framework to overcome these limitations, providing a scalable and topologically protected approach to realizing DTQWs in physical systems. 
Thouless quantum walks, or ThQWs, emerge from the interplay of two key ingredients: \textit{degenerate flat bands} and \textit{non-Abelian Thouless pumping}~\cite{PhysRevA.103.063518}, as illustrated in Fig.~\ref{fig:intro}. 

Flat-band lattices are translationally invariant tight-binding networks in which one or more Bloch bands are dispersionless—\textit{i.e.}, their energy remains constant and independent of the wavevector ${\bf k}$~\cite{leykam2018artificial,leykam2018perspective,vicencio2021photonic,danieli2024flat}. These flat bands arise from destructive interference, giving rise to eigenstates that are strictly confined in space, known as compactly localized states (CLS).

Thouless pumping, on the other hand, refers to the quantized transport of particles driven by the adiabatic and periodic modulation of a confining lattice potential~\cite{Thouless1983quantization,Citro2023thouless}. In systems with degenerate Bloch bands, this process gives rise to a non-Abelian gauge structure associated with local rotations within the degenerate subspace~\cite{PhysRevA.103.063518,Pilozzi2022}.

As we show below, these two components allow for the implementation of the basic building-blocks of DTQWs.
Specifically, the degenerate CLS replicate the coin levels while
non-Abelian pumping enables both quantized conditional transport and rotations in the coin space, analogous to the shift $\mathcal{T}$ and coin $\mathcal{R}$ operators. 
This interplay of flat bands and adiabatic transport is particularly relevant in chiral flat bands~\cite{ramachandran2017chiral}, where bipartite symmetry ensures the presence of multiple strictly flat bands at $\kappa_{0}=0$ independent of the hopping strengths.In these systems, the hopping modulation does not break the degeneracy, which is essential for the non-Abelian pumping. 

Let us discuss this mechanism in details upon  the most commonly studied DTQWs, where the walker moves along a line with site index $n$ and has two coin levels $\sigma = \pm$~\cite{nayak2000quantum}. 
In this case, the shift and coin operators are:
\bea 
\mathcal{T} &=& \sum_n  \Big[ |n+\delta_+\ra \la n| \otimes |+\ra\la+| +  |-\ra\la-| \otimes|n+\delta_-\ra \la n| \Big] \\ 
	 	\mathcal{R}& =& \sum_n  |n\ra \la n|	  \otimes\begin{pmatrix}
		\cos\theta & \sin\theta \\
		-\sin\theta & \cos\theta
	\end{pmatrix} 
\eea
where the integers $\delta_\sigma$ denote the shifts of the states $|n,\sigma\ra= |n\ra\otimes|\sigma\ra$ 
while $\theta$ creates the superposition between the two levels.

{\color{black} Implementing $ \mathcal{T} $ and $ \mathcal{R} $ by means of non-Abelian Thouless pumping
requires:  
(i) considering a lattice with $d_\nu=2$ degenerate flat bands and choosing two orthogonal compact states, $|p_n\rangle$ and $|q_n\rangle$ localized within a unit cell $n$, that play the role of coin states; and 
(ii) designing two pumping cycles $\mathcal{C}_{\mathcal{T}}$ and $\mathcal{C}_{\mathcal{R}}$ of period  $\lambda_{\mathcal{T}}$ and $\lambda_{\mathcal{R}}$ respectively, which yield conditional shift  and coin operators. 
The composition $\mathcal{C} = \mathcal{C}_{\mathcal{T}}\circ \mathcal{C}_{\mathcal{R}}$  yields one cycle of duration $\lambda=\lambda_{ \cal T}+\lambda_{\cal R}$. }

{\color{black} 
The discrete nature of the resulting ThQWs is  enforced by the topology of the lattice and the geometric structure of the pumping cycle. This process naturally generates a quantized motion made of steps connecting one unit cell to the next, each step taking one pumping period $\lambda$. Hence, the pumping period $\lambda$ represents the ThQWs time unit.  The state of the walker after $t$ steps with $t\in\mathbb{Z}$ is described by the wavefunction $|\Psi(z_t)\ra = \sum_n \left[  \psi_n^p(z_t) |p_n\rangle + \psi_n^q(z_t) |q_n\rangle \right]$ with $z_t  = z_0 + \lambda\, t$, and $|\Psi(z_{t+1})\ra= \mathcal{C} |\Psi(z_t)\ra$. 
}

The geometric properties of ThQWs manifest  in the displacement matrix $D$, which—as discussed in the Methods section and demonstrated in Ref.~\cite{PhysRevA.103.063518}—can be linked to the non-Abelian field strength. This matrix $D$ characterizes the displacement undergone  by an arbitrary superposition of the two coin states. Importantly, the trace of the displacement matrix $D$ yields the first Chern number of the degenerate band, $C_1 = \Tr[D] $, which governs the displacement of the geometric center of the final distribution via the relation $n_t = \frac{C_1 t}{2}$.

\begin{figure}[h]
	\centering
	\includegraphics [width=0.99\columnwidth]{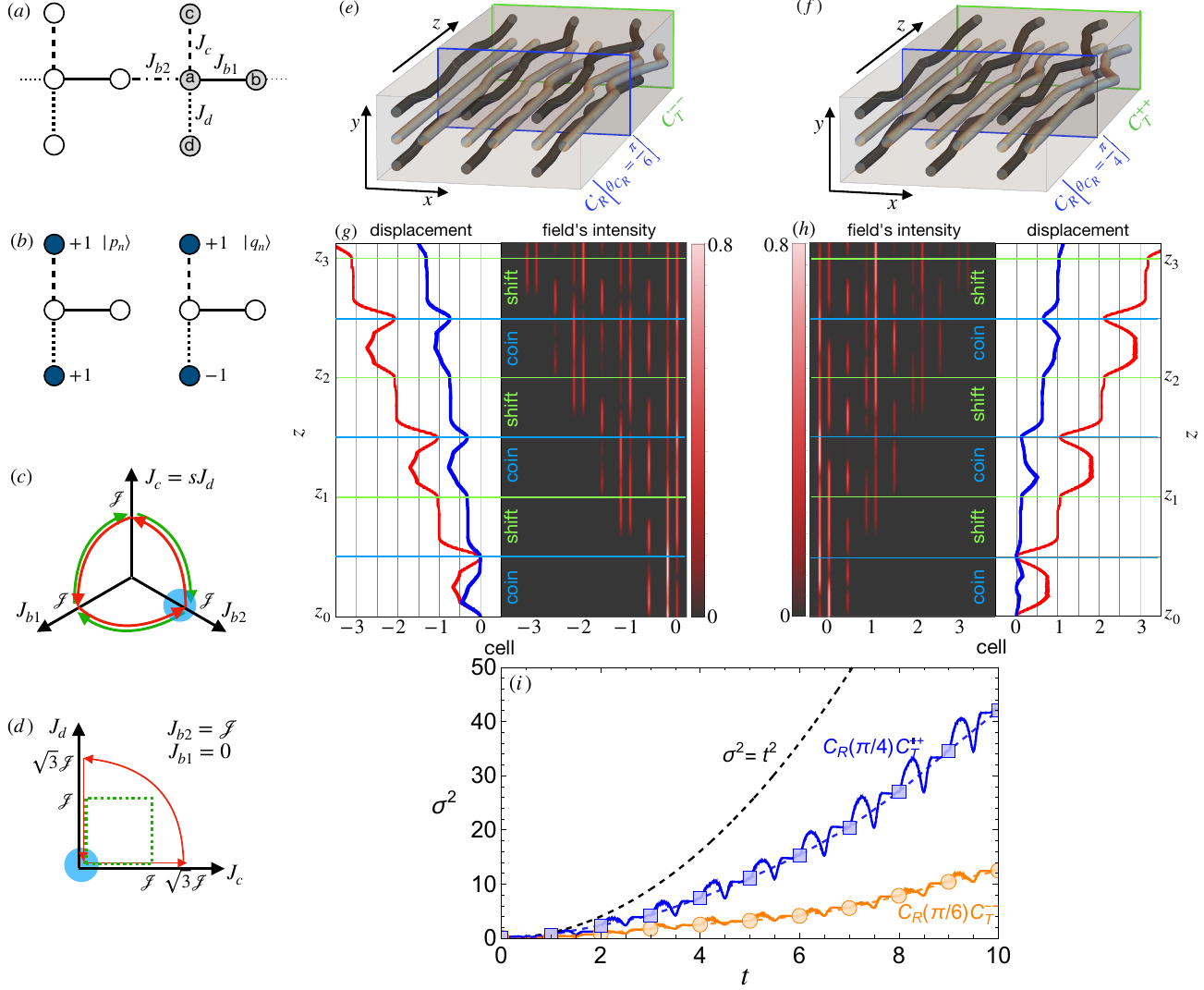}
	\caption{
		{\bf Photonic implementation of a sample uni-directional Thouless quantum walk.} 
		(a) Flat band lattice profile with the unit cell coloured in grey. 
		(b) Orthogonal symmetric $|p_n\rangle$ and antisymmetric $|q_n\rangle$ states with non-zero amplitudes coloured in dark blue.
		(c) Pumping cycles $\cyt^{+ s}$ (red) and $\cyt^{- s}$ (green) in the parameter space.  The blue circles indicate the initial point.
		(d) Same as (c) for pumping cycles $\cyr(\frac{\pi}{4})$ (red) and  $\cyr(\frac{\pi}{6})$ (green).
		(e) Illustration of three unit cells of the pumped lattice implementing a time-unit formed by a coin $\cyr(\frac{\pi}{6})$ and a shift $\cyt^{--}$. 
		(f) Same as (e) with coin $\cyr(\frac{\pi}{4})$ followed by a shift $\cyt^{++}$. 
		(g) Three steps propagation of the ThQWs generated by a step $\cyt^{--} \cyr(\frac{\pi}{6})$ from a single-cell excitation $|\Psi(z_0)\ra = \frac{1}{\sqrt{5}} |p_0\rangle + \frac{2}{\sqrt{5}} |q_0\rangle$. \textcolor{black}{In both cycles $\cyt$ in (c) and $\cyr$ in (d)  we set $\lambda \mathcal{J}= 200$}. 
		The green lines indicate the time-steps $z_t$, while the blue ones separate the coin for the shift.  
		The right panel shows the field's intensity, while the left panel shows the trace of the displacement matrix $D$ (red) and the center of mass (blue). 
		(h) Same as (g) for a ThQWs generated by a step $\cyt^{++} \cyr(\frac{\pi}{4})$. 
		{\color{black} (i) Variance $\sigma^2$ for $\cyt^{--} \cyr(\frac{\pi}{6})$ (orange) and $\cyt^{++} \cyr(\frac{\pi}{4})$ (blue) from $|\Psi(z_0)\ra = \frac{1}{\sqrt{2}} |p_0\rangle + \frac{1}{\sqrt{2}} |q_0\rangle)$. The squares and circles indicate the $\sigma^2$ at integer multiples of $\lambda$ The dashed black line guides the eye, indicating $\sigma^2=t^2$. }
	}
	\label{fig:lattice}
\end{figure}

{\color{black} We consider the implementation of ThQWs in a chain} with two degenerate flat bands~\cite{PhysRevA.103.063518}, reminiscent of the two-dimensional Lieb lattice~\cite{leykam2018artificial,leykam2018perspective,vicencio2021photonic,danieli2024flat}. The Hamiltonian features four sites {\sl per} unit cell indicated respectively as $a, b, c$ and $d$ in Fig.~\ref{fig:lattice}(a) and it reads
\be 
\small
H=\sum_n \!\lf(J_{b1}a^\dag_nb_n + J_{b2}a^\dag_nb_{n-1} + J_c a^\dag_nc_n+J_d a^\dag_nd_n+ {\rm H.c.} \rg)
\label{eq:Ham}
\ee
where $x_n^\dag$ and $x_n$ denote the creation and the annihilation operators on sites $x=a,b,c,d$ of the cell $n$. 
For  any choice of the hopping parameters, the  spectrum consists of two non-dispersive modes with longitudinal momentum $\kappa_{0}=0$ and  two dispersive modes  with longitudinal momenta $\kappa_{\pm} (k)= \pm \Delta(k)$ with $\Delta(k)=\sqrt{J_c^2+J_d^2+|J_b(k)|^2}$ and $J_{b}=J_{b1}+J_{b2} e^{ik}$. 
The chiral symmetry of the lattice~\cite{ramachandran2017chiral} implies that the Bloch states corresponding to $\kappa_0$ can be written as
\be
\begin{split}
|\phi_{1}\ra &=\frac{J_c |d_k\ra-J_d|c_k\ra}{\delta}\\
|\phi_{2}\ra &=\frac{-\rho ^2 |b_k\ra+J_{b}^*(J_c|c_k\ra+J_d|d_k\ra)}{ \rho \Delta(k)}
\end{split}
\label{eq:Bl_states}
\ee
with $\rho =\sqrt{J_c^2+J_d^2}$. 
\textcolor{black}{Let us assume that the system~\eqref{eq:Ham} is initially prepared in a Wannier state $|\Psi(z_0)\ra =\sum_{k,\ell} \alpha_{\ell} |\phi_{\ell}(k)\ra e^{i k n}$. 
An adiabatic pumping cycle $\mathcal{C}$ acting on $|\Psi(z_0)\ra$ then yields the state}
\be
|\Psi(z_1)\ra=\sum_{k\ell m}\alpha_{\ell}[W_\mathcal{C}(z_1,z_0)]_{m \ell}|\phi_{ m}(k,z_0)\ra e^{ikn}
\label{eq:evolution}
\ee 
\textcolor{black}{where  $W_\mathcal{C}(z_1,z_0)=\mathcal{P} \ \text{exp}\left[i\int_{z_0}^{z_1}\Gamma_{0}^zdz\right]$ indicates the holonomy transformation associated with the Wilczek-Zee connection $[\Gamma_z]_{\ell m}=\la\phi_{\ell}(k,z)|i\partial_z|\phi_{ m}(k,z)\ra$. Here $\mathcal{P}$ denotes the path ordering, while the indexes $m,\ell=1,2$ enumerate the basis states of the degenerate subspace and $k$ runs over the reciprocal space vectors  -- see the Methods section for details.} 

To engineer the ThQWs we initialize the lattice setting $J_c=J_d=J_{b1}=0$ and $J_{b2}=\mathcal{J}$. The Wannier states belonging to the non-dispersive modes in Eqs.~\eqref{eq:Bl_states} can be then cast as   
\be
\small	
|q_n\rangle=\frac{|c_n\ra - |d_n\ra}{\sqrt{2}} \qquad\quad
|p_n\rangle=\frac{|c_n\ra + |d_n\ra}{\sqrt{2}} 
\label{eq:states}
\ee
{\color{black} shown in Fig.~\ref{fig:lattice}(b).} These states act as the two coin levels. Following Ref.~\cite{PhysRevA.103.063518}, and as discussed in more detail in~\cite{Supple},  the pumping cycles implementing conditional shift and coin operators on these states can be straightforwardly designed. As shown in Fig.~\ref{fig:lattice}(c), the conditional shift operator cycles, denoted as $\cyt^{\xi s}$, are represented by spherical triangles on the hyperplane $J_c= s J_d$ with $s=\pm$, having anti-clockwise ($\xi=+1$) or clockwise ($\xi=-1$) orientation. In contrast,  the coin operator cycle $\cyr$, shown in Fig.\ref{fig:lattice}(d),  lies on the plane $\{J_{b2} = \mathcal{J} , \,J_{b1} = 0\}$.  The corresponding holonomic transformations in $k$-space are
\be 
\small
W_{\cyt^{\xi s}}=e^{ i\xi\frac{k}{2}\lf(\sigma_0 - s \sigma_z\rg)}\qquad 
W_{\cyr}=e^{i\theta ( \sin k\sigma_x+\cos k\sigma_y )}
\label{eq:holonomies}
\ee	
with $\sigma_i$ and $\sigma_0$ denoting the Pauli matrices and the identity. %
The above equations show that ThQWs have two important features:  (i) they yield conditional shifts determined by the topology of the driven bands; and (ii) they offer the possibility to control the shift's direction by controlling the orientation of the pumping cycles, related to the index $\xi$. 
Moreover, in the specific example discussed here,  they allow us to select which of the coin states, $| p_{n}\ra$  or   $|q_n\ra$, moves by changing the relative phase of $J_c$ and $J_d$.
Specifically, they implement the following operators:
\bea 
\cyt^{\xi+}&:& \sum_n  \Big[ |p_{n+\xi}\ra \la p_{n}|  +  |q_n\ra\la q_n| \Big] \\ \cyt^{\xi-}  &:& \sum_n  \Big[ |p_{n}\ra \la p_{n}|  +  |q_{n+\xi}\ra\la q_n| \Big] 
\eea
which respectively result in right- and left- oriented uni-directional conditional shift operators.
On the other hand, the cycle $\cyr$ yields zero net displacement while performing a rotation by an angle $\theta$ which is dependent on its precise shape~\cite{PhysRevA.103.063518}. The angle $\theta$ induces the superposition between the two flat bands.
For example, the cycle in Fig.~\ref{fig:lattice}(d) corresponds to $\theta = \frac{\pi}{4}$. In this case, reversing the orientation of the pumping cycles reverses the rotation. Composing the cycles  $ \cyt^{\xi s}$ and   $\cyr$ enables the realization of directional ThQWs with chirality $\chi = \xi s$.

To demonstrate this, in Figures~\ref{fig:lattice}(e-h) we show two examples of ThQWs corresponding, respectively, to ${\cal C}=\cyt^{--}\cyr$ with $\theta=\pi/6$ (Fig.~\ref{fig:lattice}(e,g)) and ${\cal C}=\cyt^{++}\cyr$  with  $\theta=\pi/4$ (Fig.~\ref{fig:lattice}(f,h)).
{\color{black} In Figs.~\ref{fig:lattice}(e,f) we show the schematic of three unit cells of the pumped lattice Eq.~\eqref{eq:Ham} where the blue lines mark the change between coin $\cyr$ and shift $\cyt^{\xi s}$ and the green line mark the time-steps $z_t$.} In Figs.~\ref{fig:lattice}(g,h) we display the field intensity along the waveguides for a single unit cell initial condition of the form $|\Psi(z_0)\ra = \frac{1}{\sqrt{5}} |p_0\rangle + \frac{2}{\sqrt{5}} |q_0\rangle$.  
\textcolor{black}{In our simulations, we set  $\lambda  \mathcal{J} = 200$, compatible with the parameters choices done in photonics implementations~\cite{Yan2024,Chen2025} -- see Methods for details}. 
As one can see, ThQWs enable a remarkable control on field's propagation. The left and right panels of Figs.~\ref{fig:lattice}(g,h) display the displacement of the center of mass of the distribution (blue curve) and the trace of the displacement matrix $D$ (red curve). As expected, only the latter is perfectly quantized, as it is directly related to the first Chern number -- see Methods for details.

\textcolor{black}{The variance $\sigma^2$ of a discrete time quantum walk distribution is a key indicator of its spreading behavior compared to a classical random walk. Crucially, while the variance of a classical random walk grows linearly $\sigma^2\sim t$ with the number of steps $t$, a quantum walk exhibits quadratic growth, $\sigma^2 \sim t^2$~\cite{doi:10.1126/sciadv.aat3174}. 
Fig.~\ref{fig:lattice}(i) illustrates this ballistic spreading, showing the variance $\sigma^2$ for two different ThQWs: one defined by the time-step $\cyt^{--} \cyr(\frac{\pi}{6})$ (orange) and another by $\cyt^{++} \cyr(\frac{\pi}{4})$ (blue), analogous to Figs.~\ref{fig:lattice}(e,f). Both ThQWs start from a symmetric single-cell excitation $|\Psi(z_0)\ra = \frac{1}{\sqrt{2}} |p_0\rangle + \frac{1}{\sqrt{2}} |q_0\rangle)$ and evolve for ten time-steps. The solid curves indicate the variance $\sigma^2$ during the evolution, while squares and circles  indicate the value of $\sigma^2$ at the multiples of $\lambda$.
As shown, the coin parameter $\theta$ significantly influences the walk’s dynamics by modulating the effective spreading rate~\cite{doi:10.1126/sciadv.aat3174}. Indeed, the variance follows $\sigma^2 \simeq f t^2 $ where the prefactor $f$ depends on the coin angle $\theta$ as well as on the initial state. By tuning $\theta$, one can tailor the quantum walk for the desired variance --  a property that can be exploited to enhance the efficiency of quantum search and sampling algorithms. 
In ThQWs, this tuning can be done by properly designing the pumping cycle $\cyr$ that implements the coin operator.}

\begin{figure}[t]
	\centering
	\includegraphics [width=0.75\columnwidth]{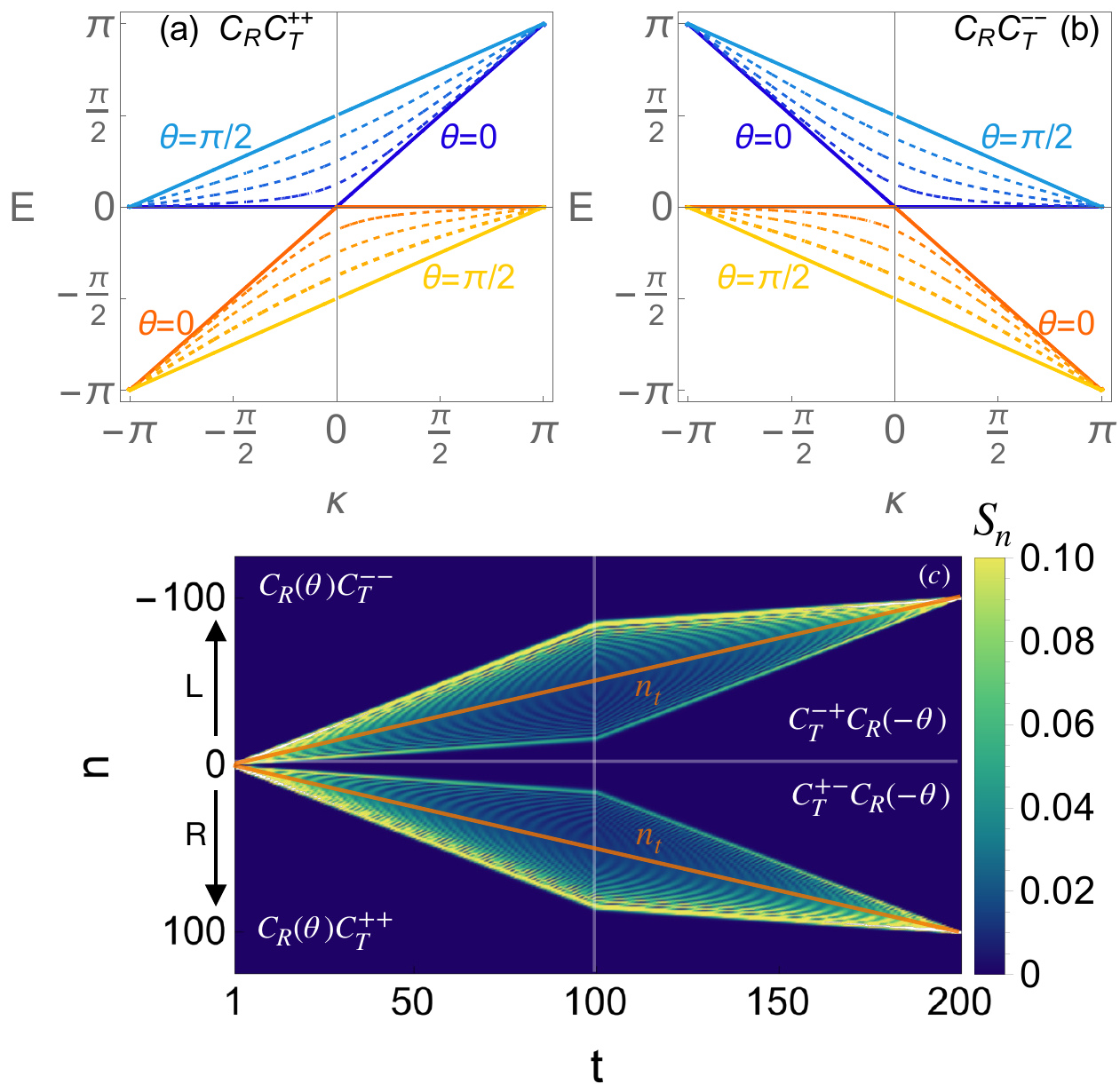}
	\caption{
		{\bf Parity breaking in Thouless quantum walk.} 
		Floquet quasi-energies for the quantum walks $C_{\mathcal{R}} C_{\mathcal{T}}^{++}$ (a)  and $C_{\mathcal{R}} C_{\mathcal{T}}^{--}$ (b) for different values of the coin angle $\theta \in[0,\pi/2]$
		(c) Example of propagation along composite cycles starting from a symmetric single-cell excitation $|\Psi(z_0)\ra = ( |p_0\rangle + |q_0\rangle)/\sqrt{2}$. The pattern is generated by means of a two stage procedure. In the first stage we perform the $C_{\mathcal{R}} C_{\mathcal{T}}^{++}$ for $n\geq 0$ and $C_{\mathcal{R}} C_{\mathcal{T}}^{--}$ for $n\leq 0$  and suitably overlapping the two cycles for $n=0$. In the second stage which start at $t=100\lambda$, we apply time-reversal symmetry and we exchange $|p_n\rangle$ and $|q_n\rangle$ while keeping the chirality $\chi$ of the walks. 
	}
	\label{fig:comp-pat}
\end{figure}

To illustrate the parity-broken nature of ThQWs, we analyze the dynamic structure of the quantum walk equations in the continuous time limit, as described in Ref.~\cite{Chandrashekar2010relationship}. By doing so, for the quantum walk generated by the cycles $\cyt^{\xi s} \cyr(\theta)$,  we obtain
\begin{equation}
\begin{split}
\partial_t \Psi
=& - \xi \left[  \cos\theta  \frac{\sigma_0 + s \sigma_z}{2} + \sin\theta \frac{ i \sigma_y  + s \sigma_x }{2}  \right]  \partial_n \Psi +\\
&+ \left[  (\cos\theta-1) \mathbb{I} + i \sin\theta\, \sigma_y \right] \Psi	\\
\end{split}
\label{eq:state_d}
\end{equation}
with $\Psi=( \psi_{n}^p (t), \psi_{n}^q (t))^T$ -- see~\cite{Supple} for details.  
For $\xi=\pm1$ the above equation accounts for a right-moving(+) or left-moving(-) ThQWs. 
In the limit $\theta=0$, for each value of $\xi$ Eq.\eqref{eq:state_d} describes a right-handed $(\chi = +1)$ and a left-handed $(\chi = -1)$ Weyl particle. 
Equivalently, the two Floquet quasi-energies $E$~\cite{Supple,kitagawa2010} associated with the processes $C_{\cal R}C^{++}_{\cal T}$ and  $C_{\cal R}C^{--}_{\cal T}$  produce two anisotropic Floquet bands whose structure depends on the angle $\theta$, as shown  in Fig.~\ref{fig:comp-pat}(a) and Fig.~\ref{fig:comp-pat}(b). For $\theta = 0$, the system exhibits a  flat Floquet band $E=0$ and a linearly dispersive Floquet band $E=\xi k$ which are respectively associated with $|q_n\rangle$ and  $|p_n\rangle$ in the cycle $C_{\mathcal{R}} C_{\mathcal{T}}^{++} $ and the reverse for the cycle $C_{\mathcal{R}} C_{\mathcal{T}}^{--}$. 
For $\theta \neq 0$, the states $|p_n\rangle$ and $|q_n\rangle$ become coupled, yet the left-right directionality is preserved.  
Linear Floquet band dispersion $E=\frac{\xi}{2} (k\pm \pi)$ is restored for $\theta=\pi/2$ and the two states move in unison with half the group velocity of a $\theta=0$ single state. 

By composing ThQWs with different parities, more complex patterns can be generated. For instance, the pattern in Fig.~\ref{fig:comp-pat}(c) is produced through a two-stage procedure in which the ThQWs are combined with their counterparts obtained via the time-reversal operator and a chirality-preserving spin-inversion operator.
In the first stage, the cycles $C_{\mathcal{R}} C_{\mathcal{T}}^{++} $ (for $n \geq 0 $) and $C_{\mathcal{R}} C_{\mathcal{T}}^{--} $ (for $n \leq 0 $) are performed, with an appropriate overlap at $n = 0 $ to ensure smooth connectivity. In the second stage, which begins after $100$ steps, time-reversal symmetry is applied, and the roles of $|p_n\rangle$ and $|q_n\rangle$ are exchanged while preserving the chirality $\chi$ of the time-reversed walks. This therefore involves combining ThQWs generated by the cycles $C_{\mathcal{R}}(\theta) C_{\mathcal{T}}^{++} $ and $C_{\mathcal{R}} (\theta) C_{\mathcal{T}}^{--} $ with those generated by $C_{\mathcal{T}}^{+-} C_{\mathcal{R}} (-\theta)$ and $C_{\mathcal{T}}^{-+} C_{\mathcal{R}}(-\theta) $, respectively. The whole process maps the initial state $(|p_0\ra+|q_0\ra)/\sqrt{2}$ onto the state $(|q_{-t/2}\ra+|p_{t/2}\ra)/\sqrt{2}$ therefore generating quantum correlations between spin and position coordinates. 
 Also in this case, the displacement of the middle point of the two distributions is governed by  topology. 

The high degree of control offered by ThQWs enables the realization of different classes of quantum walks.
For example, varying $\theta$ periodically in the cycle $\cyr$ at every step $ z_t$ yields Floquet quantum walks~\cite{katayama2020floquet}, while varying $\theta$ with $n$ spatially modulates ThQWs, yielding topologically nontrivial states -- {\it e.g.} by designing a split-step advancement operator $\mathcal{C} \equiv \cyt^{--} \cyr(\theta(n)) \cyt^{++} \cyr(\theta(n))$ as discussed in Ref.~\cite{Kitagawa2009universal}. 

\begin{figure}[t]
	\centering
	\includegraphics [width=0.95\columnwidth]{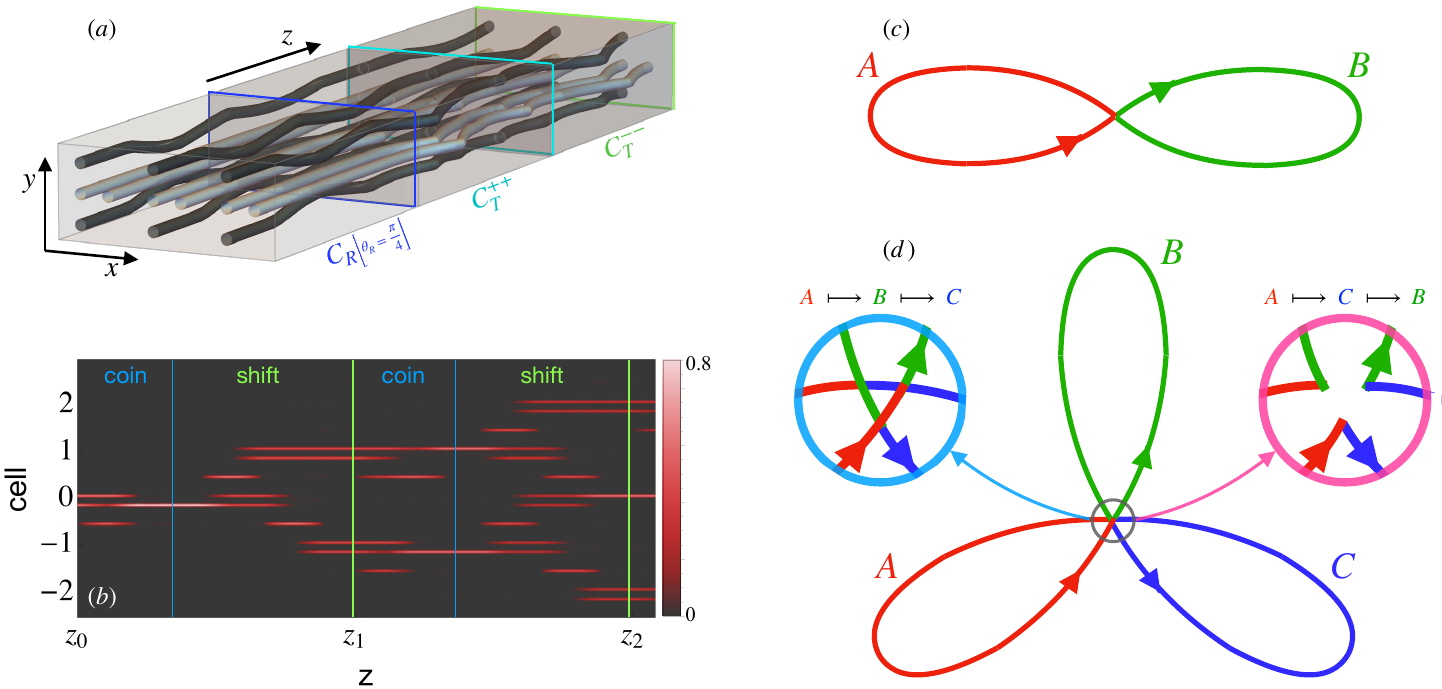}
	\caption{
		{\bf Photonic implementation of bi-directional Thouless quantum walk.} 
		(a) Illustration of three unit cells of the lattice pumped by cycle $\cyr$ with $\theta=\pi/4$ (marked by blue lines), cycle $\cyt^{++}$ (marked by cyan lines) and cycle $\cyt^{--}$ (marked by green lines). 
		(b) Two steps propagation of the walk from a single-cell excitation. The green lines indicate the time-units $ z_s$, while the blue ones separate the coin $\cyr$ and shift $\cyt^{\xi s}$. 
		(c) Schematic illustrations of the two-lobes curve in the parameter space representing a two-cycle walk $W_M(A\cdot B)$. The red and green lobes represent cycle $A$ and $B$. (d) Three lobes curve representing the three-cycle walk $W_M(A\cdot B\cdot C)$. The red, green and blue lobes represent cycle $A$, $B$ and $C$ respectively. The two zooms schematically indicate the cycle order $A\rightarrow B \rightarrow C$ (blue) and $A\rightarrow C \rightarrow B$ (magenta) around the junction point of the three cycles.
	}
	\label{fig:standardQW}
\end{figure}

{\color{black} 
The pumping cycles $\cyt^{\xi s}$ in Eq.~\eqref{eq:holonomies} can be used to generate ThQWs with restored parity symmetry. For instance, composing cycles $\cyt^{\xi +}$ and $\cyt^{\bar{\xi}-}$ for $\bar{\xi} = -\xi$ yields the shift operators
\bea 
\cyt^\xi= \cyt^{\xi+} \circ \cyt^{\bar{\xi} -}&:&  \sum_n  \Big[ |p_{n+\xi}\ra \la p_{n}|  +  |q_{n-\xi}\ra\la q_n| \Big] 
\eea
In Fig.~\ref{fig:standardQW}(a) we show three unit cells of pumped lattice waveguides, where this shift is composed with a coin $\cyr$ with $ \theta=\pi/4$.
Fig.~\ref{fig:standardQW}(b) then shows the field intensity along the waveguides resulting from a single-cell excitation with $|\Psi(z_0)\ra = \frac{1}{\sqrt{5}} |p_0\rangle + \frac{2}{\sqrt{5}} |q_0\rangle$. The transitions from the coin operator $\cyr$ to the shift operator $\cyt$ are marked by the blue lines, while the time-units $z_1$ and $z_2$ are marked by the green lines. }

The freedom to combine and engineer different pumping cycles offers the possibility to probe the underlying non-Abelian gauge structure. Due to the non-Abelian multiband nature of the adiabatic evolution, the final probability distribution generated by the quantum walk depends on the order in which the shift and rotation cycles -- denoted as $\mathcal{C}_T$ and $\mathcal{C}_R$ -- are applied. 
In the asymptotic limit of many steps, swapping the order of these cycles effectively corresponds to initiating the walk from a different quantum state, which can significantly alter the resulting distribution.

More striking signatures of the geometric and topological character of ThQWs emerge when unit steps are composed of at least three non-commuting pumping cycles. In such cases, only cyclic permutations of the cycle sequence yield quantum walks that are asymptotically equivalent -- up to a change in the initial state -- while other permutations result in fundamentally distinct quantum walks.
This behavior is illustrated in Fig.~\ref{fig:standardQW}(c,d). For a generic two-cycle quantum walk $W(A\cdot B)$, the trajectory traces $M$ times a two-lobe curve in the parameter space shown in Fig.~\ref{fig:standardQW}(c) whose topology remains invariant under permutation of $A$ and $B$. In contrast, for a three-cycle quantum walk $W_M(A\cdot B\cdot C)$, the trajectory forms a three-lobe curve in the parameter space. Changing the cycle order leads to inequivalent walks, each generating qualitatively different dynamical behavior – see~\cite{Supple} for more details. In Fig.~\ref{fig:standardQW}(d) the two possible orders
are shown in the two zooms around the intersection point – namely,  $A\rightarrow B \rightarrow C$
(blue left zoom) and $A\rightarrow C \rightarrow B$ (magenta right zoom).

Considering one-dimensional lattices with $d_\nu \geq 3$ degenerate flat bands may allow for the implementation of ThQWs with three or more coin levels. 
One way could be extend the tripod uni-cell shown in Fig.~\ref{fig:lattice}(a) to a $M$-pod unit cell and build a lattice that respects the bipartite symmetry~\cite{ramachandran2017chiral}. This symmetry relation guarantees $M-2$ flat bands at $\kappa_0 = 0$ energy. 
This implementation of ThQWs through chiral flat band lattices is not restricted to DTQWs on a line, but could be exploited to design ThQWs on general graphs.

\section{Discussion}\label{sec3}

Non-Abelian holonomies and geometric phases have emerged as powerful tools in theoretical physics, particularly within gauge theories, topological phases of matter, and quantum computation.
In this work, we show that non-Abelian Thouless pumping provides a  framework for realizing quantum walks with coin and shift operators defined through  non-Abelian holonomies. Specifically, we demonstrate that in a lattice with multiple degenerate flat bands, Thouless pumping enables the implementation of holonomic quantum gates that act on the spatial degrees of freedom of a quantum particle and entangle them with its internal coin states. The topological origin of the pumping ensures quantized transport and can be naturally related to a class of discrete time quantum walks that we called Thouless Quantum Walks, or ThQWs.

ThQWs allow for a high-degree of control and tunability. Specifically, we show that they can be engineered to \textcolor{black}{selectively} break parity or time-reversal symmetry yielding a  directionally chiral evolution that is effectively described by Weyl-like equations in the continuum limit—mirroring relativistic dynamics but in a lattice-based, topologically protected framework. 
The topological  and geometric nature of Thouless pumping endows ThQWs with robustness against  certain kinds of noise and dynamical perturbations, a highly desirable property for quantum technologies.
Moreover, we reveal that the combination of parity and time-reversal symmetry breaking allows for the emergence of well-defined quantum correlations in the long-time limit. This feature can be used to control entanglement patterns and engineer specific correlation structures, potentially useful for quantum communication protocols or quantum simulations.

Our approach not only opens a novel direction for the application of non-Abelian holonomies in dynamic quantum systems but also lays the groundwork for extending these ideas beyond the one-dimensional case studied here. Future generalizations may include higher-dimensional quantum walks, which could underlie lattices with higher Chern numbers, or systems with larger internal (coin) spaces, thus enabling more complex forms of quantum information encoding and manipulation.

\section{Methods} 

\subsection{Geometric properties of ThQWs}
The displacement generated in a quantum walk for the system initially prepared in an arbitrary superposition of two CLSs   centered on site $n$ belonging to the level $\kappa_\nu$ is given by 
\be 
\Delta x=\sum_{m\ell}\alpha^*_m \alpha_\ell D^\nu_{m\ell},
\label{eq:displ}
\ee
where the coefficients $\alpha_\ell$ define the initial state 
\be
|\Psi(z_0)\rangle =  \sum_{k,\ell} \alpha_{\ell} |\phi_{\ell}(k)\ra e^{i k n},
\label{eq:Wstate}
\ee
\textcolor{black}{This Wannier state is defined for the normalized Bloch states $|\phi_{\ell}(k)\ra$. In this sum, the coefficients $\alpha_\ell$ define the superposition of the Bloch states of the $E=0$ degenerate subspace for the index $\ell=1,2$, while $k$ runs over the reciprocal space.
After an adiabatic pumping cycle $\mathcal{C}$, the state $|\Psi(z_0)\ra $ in Eq.~\eqref{eq:Wstate} turns to}
\be
\color{black}{|\Psi(z_1)\ra=\sum_{k\ell m}\alpha_{\ell}[W_\mathcal{C}(z_1,z_0)]_{m \ell}|\phi_{ m}(k,z_0)\ra e^{ikn}}
\label{eq:evolution_meth}
\ee 
\textcolor{black}{where  $W_\mathcal{C}(z_1,z_0)=\mathcal{P} \ \text{exp}\left[i\int_{z_0}^{z_1}\Gamma_{0}^zdz\right]$ indicates the holonomy transformation associated with the Wilczek-Zee connection $[\Gamma_z]_{\ell m}=\la\phi_{\ell}(k,z)|i\partial_z|\phi_{ m}(k,z)\ra$. Here $\mathcal{P}$ denotes the path ordering, while the indexes $m,\ell=1,2$ enumerate the degenerate basis states.}  
The displacement matrix $D$ is a square matrix which can be written as
\be
[D]_{m\ell}=\frac{1}{2\pi} \int_{z_0}^{z_1}\!\!\!dz \int_{-\pi}^{ \pi}  \!\! \!\!  dk \lf[
W^\dag_\mathcal{C}\,{\cal F}_{k z}W_\mathcal{C}\ \rg]_{m\ell} 
\label{eq:displ-matr}
\ee
with ${\cal F}_{kz}=\partial_k \Gamma_z - \partial_z \Gamma_k + i\lf[\Gamma_z, \Gamma_k\rg]$ denoting the non-Abelian field strength matrix. 
The above expression directly shows the geometric and topological properties of the displacement generated in quantum walks and relates the properties of the walk to the Wilczek-Zee\cite{PhysRevLett.52.2111} connection along $z$  and $k$ defined as 
\be
\begin{split}
[\Gamma_z]_{\ell m} &=\la\phi_{\ell}(k,z)|i\partial_z|\phi_{ m}(k,z)\ra 
\end{split}
\label{eq:WZ_conn_z}
\ee 
and
\be
[\Gamma_k]_{\ell m}=\la\phi_{\ell}(k,z)|i\partial_k|\phi_{ m}(k,z)\ra.
\label{eq:WZ_conn_k}
\ee 
Using Eq. \ref{eq:WZ_conn_z} one can explicitly calculate the holonomies generated in the pumping cycles $\cyt$ and $\cyr$.

Specifically, denoting as $W_{\cyt^{\xi s}}$  the holonomy corresponding to the cycle $\cyt^{\xi s}$  on the hyperplane $J_c= s J_d$ with $s=\pm$ with anti-clockwise ($\xi=+1$) or clockwise ($\xi=-1$) orientation shown in Fig.\ref{fig:lattice}(c), we can write
\be 
\begin{split}
W_{\cyt^{\xi s}}=e^{ i\xi\frac{k}{2}\lf(\sigma_0 - s \sigma_z\rg)}
= \begin{pmatrix}
e^{ i\xi \frac{k}{2} (1- s ) }   &0\\
0& e^{ i\xi \frac{k}{2} (1+s ) } 
\end{pmatrix} 
\end{split}
\label{eq:holonomies_C1_m}
\ee	 
In contrast, the holonomy associated to the cycle $\cyr$ defined on the plane $\{J_{b2} = J, \,J_{b1} = 0\}$ and shown in Fig.\ref{fig:lattice}(d) is
\be
\begin{split} 
W_{\cyr}&=e^{i\theta ( \sin k\sigma_x+\cos k\sigma_y )} 
= \begin{pmatrix}
\cos \theta  &e^{i k} \sin \theta\\
-e^{-i k} \sin \theta & \cos \theta   
\end{pmatrix} 
\end{split}
\label{eq:holonomies_C2_m}
\ee
where the angle $\theta$ can be expressed by the integral along the path $\cyr$ as
 \be \theta=\int_{\cyr} \frac{(J_{c}\partial_z J_{d}- J_{d}\partial_z J_{c})\sin k}{\rho^2 \Delta^2(k)}.\ee 
 
 \subsection{Implementation of ThQWs  in photonic waveguides and integrated photonics}

The implementation of  ThQWs poses several technical challenges, associated with the realization of a large number of non-Abelian holonomic transformations. 
So far, most experimental work has focused on implementations based on  laser-written glass structures such as those realized in Ref.~\cite{Yan2024} and on silicon photonics implementations~\cite{Chen2025}.
In realistic experiments based on femtosecond laser-written glass structures, the waveguides typically have dimensions on the order of $5$--$10\,\mu\mathrm{m}$, with separations ranging from $10$ to $20\,\mu\mathrm{m}$. Typical effective refractive indices and operating wavelengths in vacuum are $n_e = 1.5$ and $\lambda_0 = 0.8\,\mu\mathrm{m}$, respectively. These values lead to an effective optical potential per waveguide of approximately $V_0 \simeq 1\,\mathrm{meV}$ and a field decay length of a few micrometers. The corresponding coupling constants in this regime are typically on the order of
\[
J \simeq 10^{-3}\,\mu\mathrm{m}^{-1}.
\]
To satisfy the adiabatic condition, the modulation wavelength $\lambda$ must fulfill $\lambda J \gg 1$. For instance, setting $\lambda = 2\,\mathrm{cm}$ gives $\lambda J \sim 20$, which allows approximately 10 modulation cycles within waveguides of length $20\,\mathrm{cm}$. This length is realistic and permits the neglect of loss effects. Reducing the losses or finding strategies to  perform non-adiabatic holonomic gates would significantly enhance the maximum number of steps.  
A comprehensive discussion of non-Abelian holonomic effects, derived from coupled-mode theory in photonic waveguide lattices, can be found in Ref.~\cite{snyder1983weakly,Snyder1972coupled,Marcuse1973coupled,yariv2003coupled,Rodriguez2015quantum,Pinske2022symmetry}.

\subsection{Robustness of ThQWs }

In practical implementations of ThQWs, various imperfections and non-idealities can affect the performance of ThQWs.
Deviations from the ideal pumping cycle -- such as incomplete loops or distorted parameter trajectories -- may, in general, reduce the fidelity of holonomic gates. However, their impact on the shift and rotation cycles is qualitatively different. The topological protection inherent in Thouless pumping ensures that such imperfections introduce only exponentially suppressed corrections to the shift cycle $\cyt$, as long as the energy gap protecting the degenerate subspace remains open throughout the evolution. In contrast, the rotation cycle $\cyr$ is only geometrically protected and is therefore more sensitive to imperfections. As a result, we expect that deviations from ideal pumping cycles lead to small step-to-step fluctuations in the rotation angle $\theta$. The effect of this kind of dynamical disorder has been discussed  e.g. in Refs.\cite{Panahiyan2018controlling,mastandrea2023localization}, and it critically depends on the disorder strength. Clearly, a strong lattice disorder breaks translational symmetry and can induce localization, potentially hindering quantum walks.

Imperfect control on the Hamiltonian parameters may affect the degeneracy or the flatness of the energy bands involved in the QW. These effects introduce dynamical phase errors and wavepacket dispersion, which in turn degrades holonomic precision and coherence. To a certain extent, as discussed in Refs.\cite{PhysRevA.48.1687,danieli2024nonabelian}, these effects can be controlled by appropriately choosing the duration of the cycle in such a way that the pumping process does not resolve the small energy splitting introduced by weak control of the Hamiltonian parameters.
Loss mechanisms such as photon decay or dephasing introduce non-unitarity, reduce the visibility of topological transport and essentially limit the number of steps in quantum walks, which creates a trade-off with the adiabaticity condition.  Furthermore, non-adiabatic effects arising from rapid parameter modulation lead to interband transitions and incomplete holonomies, undermining both the quantization of transport and the desired entanglement patterns.
Despite these challenges, the topological nature of Thouless pumping and the geometric robustness of holonomies offer resilience to moderate imperfections. Nevertheless, achieving high-fidelity control in THQWs requires careful mitigation of non-idealities through optimized cycle design, flat-band engineering, and coherence-preserving platforms.

\begin{center}
\begin{tabular}{ | m{4 cm} | m{4 cm}| m{4 cm} | } 
\hline
\hline
\textbf{Non-ideality} &  \textbf{Effect}& \textbf{Model and correction strategy}\\
\hline
\hline
Cycle fabrication inaccuracies  & \begin{itemize} \item Deviation from flat band condition  \item Imperfect or non-adiabatic cycles \end{itemize} & Identification  and optimization of robust cycles\\
\hline
Refractive index inhomogeneity, disorder  & \begin{itemize}\item  Transitions between eigenstates \item Reduced $\cyr$ fidelity \end{itemize} & Accurate cycle design to reduce the impact of disorder  \\
\hline
Imperfect preparation of the initial state &  \begin{itemize} \item Dispersive propagation \item Apparent gate infidelity\end{itemize} & Optimization of the initial state \\
\hline
Photon losses, wavelegth and inter-waveguide crosstalk   & \begin{itemize} \item Reduced signal intensity and coherence\item Limit on ThQWs number of steps \item  Unintended mode mixing\end{itemize} & \begin{itemize} \item Optimal control to design non-adiabatic holonomic gates 
\item Modeling using dissipative methods \end{itemize}\\  \hline
\end{tabular}
\end{center}

\section{Data Availability} 
The data provided in the manuscript are available upon request.

\section*{Acknowledgment}
The Authors acknowledge fruitful discussions with A. Coppo and A. Petri.
This work was co-funded by European Union - PON Ricerca e Innovazione 2014-2020 FESR /FSC - Project ARS01$\_00734$ QUANCOM, Project PNRR MUR PE$\_0000023$-NQSTI and PNRR MUR project CN 00000013-ICSC.

\bibliography{rmd7}

\end{document}